\documentclass[aps,prb,twocolumn,superscriptaddress,showpacs]{revtex4}
\usepackage{amsmath}	% For {align}
\usepackage{amssymb}	% Maybe unnecessary... \hslash
\usepackage{times}	% To change from Computer Modern to Times font
\usepackage{mathptm}	% To get maths in Times font
\usepackage{graphicx}	% For the figures

\begin{document}

%\preprint{}

\title{Experimental Investigation of Microwave Enhanced Cotunneling in SET Transistors}


%\author{Mikkel Ejrnaes}
%\author{M. Ejrn\ae s}
%\email[Corresponding author. e-mail: ]{ejrnaes@fisica.cib.na.cnr.it}
%\altaffiliation[Present address: ]{Superconductivity Department, C.N.R.-Institute of Cybernetics, Comprensorio Olivetti, I-80078 Pozzouli, Napoli, Italy.}

\author{Martin Manscher}
\email[Corresponding author. e-mail: ]{manscher@fysik.dtu.dk}
\affiliation{Department of Physics, B309, Technical University of Denmark, DK-2800 Lyngby, Denmark}

\author{Marko T. Savolainen}
\affiliation{Department of Physics, University of Jyv\"askyl\"a, FIN-40351 Jyv\"askyl\"a, Finland.}

\author{Jesper Mygind}
\affiliation{Department of Physics, B309, Technical University of Denmark, DK-2800 Lyngby, Denmark}


\date{\today}

\begin{abstract}
Cotunneling is an important error process in the application of single electron tunneling devices for metrological and electronic applications. Here we present an experimental investigation of the theory for adiabatic enhancement of cotunneling by coherent microwaves. The dependence is investigated as function of temperature, gate voltage, frequency, and applied microwave power. At low temperatures and applied power levels, the results are consistent with theory, using only the unknown damping in the microwave line as a free parameter. However, the results indicate that the effects of temperature, frequency and microwave power are not independent, contrary to what is suggested by theory.
\end{abstract}

% Pasquier et al.: 73.20.Dx, 72.20.Jv, 73.40.Gk
% Eiles et al.: 73.40.Gk, 73.40.Rw
% Keller et el.: 73.23.Hk, 85.30.Wx
% Covington et al.: 73.40.Gk, 73.23.Hk, 85.30.Wx
% Flensberg: 73.23.Hk, 73.61.-r

%72.20.Jv  Charge carriers: generation, recombination, lifetime, and trapping
%73.20.Dx  (does not exist)
%73.23.Hk  Coulomb blockade; single-electron tunneling  
%73.40.Gk  Tunneling (for tunneling in quantum Hall effects, see 73.43.Jn)  
%73.40.Rw  Metal-insulator-metal structures  
%73.61.-r  Electrical properties of specific thin films
%85.30.Wx  (does not exist)
%85.35.-p  Nanoelectronic devices


\maketitle

\section{Introdution\label{sec:intro}}
The unique properties of single electron tunneling (SET) devices has made them the subject of extensive research. This research includes applications such as current standards,\cite{Kautz} capacitance standards,\cite{Keller} electrometers,\cite{Aassime} quantum computing,\cite{Vion} and thermometry.\cite{Kauppinen,Bergsten} The first-order behaviour is adequately described by the well-established {\em orthodox theory}.\cite{AverinLikharev,IngoldNazarov} The so-called {\em cotunneling}, a second-order phenomenon in which one electron tunnels through each of the SET junctions at the same time, has also received some attention, as this constitutes an important error process in many SET applications. The dependence of the cotunneling current on voltage and temperature has been studied theoretically \cite{AverinNazarov} as well as experimentally.\cite{Geerlings,Eiles,Pasquier,Paul} Furthermore, Covington et {\em al.}\cite{Covington} have studied the frequency dependence for 4- and 6-junction pumps. This paper investigates experimentally the theoretical prediction by Flensberg \cite{Flensberg} that the cotunneling current should depend not only on temperature and voltage, but also on the amplitude and frequency of an applied oscillating field.

The paper is organized as follows: The theory for cotunneling derived by Flensberg and others is briefly summarized in section \ref{sec:theory}. The measurement setup is thoroughly described in section \ref{sec:setup}. The experimental results on cotunneling are presented in section \ref{sec:results}, and in section \ref{sec:discussion} the results are discussed. Finally, section \ref{sec:conclusion} summarizes our conclusions.

\section{Theory\label{sec:theory}}

{\em Cotunneling} (also referred to as {\em macroscopic quantum tunneling of electric charge} or \mbox{\em q-MQT}) is a second-order process by which an electron tunnels through the junctions of the SET transistor via an intermediate virtual state.\cite{AverinOdintsov,AverinNazarov} % The two tunneling events in cotunneling have to occur within the lifetime of the intermediate state, which is about $\hslash/\Delta E$, where $\Delta E$ is the width of the energy level.{\bf Find a reference}
The inelastic cotunneling current in a single SET transistor for low temperature ($k_\mathrm{B} T_\mathrm{e} \ll \Delta^\pm$) and bias voltage ($eV_\mathrm{DC} \ll \Delta^\pm$) was derived by Averin and Nazarov as\cite{AverinNazarov}
\begin{equation}\label{eq:DCcotun}
I_\mathrm{cot}^\mathrm{(3)} = \frac{R_\mathrm{K}}{24 \pi^2 R_1 R_2} \left( \frac{1}{\Delta^+} + \frac{1}{\Delta^-} \right)^2
 \left[ \left(2 \pi k_\mathrm{B} T_\mathrm{e}\right)^2 + \left(eV_\mathrm{DC}\right)^2\right] V_\mathrm{DC}
\end{equation}%this is established!!!!!!!?
Here, $R_{1,2}$ are the tunneling resistances of the left and right electrodes, respectively, $\Delta^\pm = (e/C_\Sigma)(e/2 \mp C_\mathrm{G} V_\mathrm{G}) \; (\mathrm{mod}\;(e^2/C_\Sigma))$ are the energies to add/remove one electron to/from the island, $T_\mathrm{e}$ is the temperature of the electron system, and $R_\mathrm{K} = h/e^2$. In the maximum blockade state, $\Delta^\pm$ reduce to the charging energy $E_\mathrm{C} = e^2/2C_\Sigma$, where $C_\Sigma = C_1 + C_2 + C_\mathrm{G}$ is the total capacitance of the island (the ground capacitance is assumed negligible). The validity of Eq. \ref{eq:DCcotun} has been verified experimentally in metallic systems by several researchers: Geerlings et {\em al.}\cite{Geerlings} first reported the observation of cotunneling current and its scaling with the conductances and voltage. The scaling of the current, rather than the quantitative value, made the results clearly distinguishable from thermally enhanced sequential tunneling. Later, the experiments by Eiles et {\em al.}\cite{Eiles} confirmed the temperature and gate bias dependence quantitatively. Furthermore, Pasquier et {\em al.}\cite{Pasquier} and other groups have investigated cotunneling in 2DEG systems. Since it seems well-confirmed by experiments, Eq. \ref{eq:DCcotun} will be assumed valid here.

Flensberg \cite{Flensberg} has extended the analysis by Averin \& Nazarov to the case where a harmonically varying signal $V_\mathrm{AC} \cos (2\pi f t)$ is applied on top of the DC bias. His result for the cotunneling current, obtained by expansion to the third order in the energies, becomes in the adiabatic (low-frequency, low-temperature, low-amplitude) limit
\begin{multline}\label{eq:ACcotun3rdOrder}
I_\mathrm{cot}^\mathrm{(3)} = \frac{R_\mathrm{K}}{24 \pi^2 R_1 R_2} \left( \frac{1}{\Delta^+} + \frac{1}{\Delta^-} \right)^2 \\
\times \left[ \left(2 \pi k_\mathrm{B} T_\mathrm{e}\right)^2 + \left(eV_\mathrm{DC}\right)^2 + \tfrac{3}{2} \left(eV_\mathrm{AC}\right)^2 \right] V_\mathrm{DC}
\end{multline}
In restating his equations we assume (as is the case for our experiments) that the alternating bias is applied to the left lead, while the right is kept at a constant potential.

Eq. \ref{eq:ACcotun3rdOrder} is not surprising, as this is what one would obtain from Eq. \ref{eq:DCcotun} by making the substitution $V_\mathrm{DC} \rightarrow V_\mathrm{DC} + V_\mathrm{AC} \cos (2\pi f t)$ and averaging over time. However, because of the power expansion approach, Flensberg was able to derive also the next leading order correction to Eq. \ref{eq:ACcotun3rdOrder} by an expansion to fourth order in the energies:
\begin{multline}\label{eq:ACcotun4thOrder}
I_\mathrm{cot}^\mathrm{(4)} = \frac{R_\mathrm{K}}{48 \pi^2 R_1 R_2} \left( \frac{1}{\Delta^+} + \frac{1}{\Delta^-}\right) ^2 \left( \frac{1}{\Delta^-} - \frac{1}{\Delta^+}\right) \\
\times \left[ \left(2 \pi k_\mathrm{B} T_\mathrm{e}\right)^2 + \tfrac{3}{4} \left(eV_\mathrm{AC}\right)^2 + \left(h f\right)^2 \right] \left(eV_\mathrm{AC}\right)^2
\end{multline}
The temperature, amplitude and frequency are assumed low enough that an expansion in powers of these is appropriate. Note that in the case of maximum Coulomb blockade, $I_\mathrm{cot}^\mathrm{(4)}$ vanishes, since then $\Delta^\pm = E_\mathrm{C}$ and thus the third multiplicative term becomes zero.

The aim of the work on cotunneling presented here is to experimentally verify or disprove Eq. \ref{eq:ACcotun3rdOrder}. We do this by applying microwaves to the device and measuring the differential conductance by a lock-in technique in zero DC bias. This ensures that the current predicted by Eq. \ref{eq:ACcotun4thOrder} does not contribute to the result. Measurements to verify Eq. \ref{eq:ACcotun4thOrder} are planned for the near future.

\begin{figure}
\resizebox{\columnwidth}{!}{\includegraphics{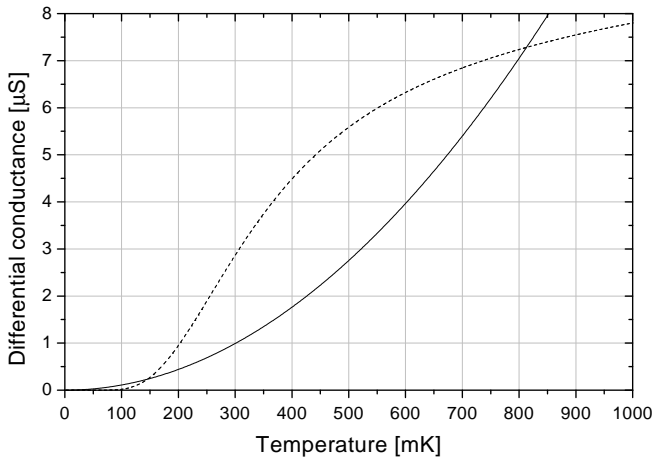}}%
\caption{\label{fig:Simulation} The orthodox theory (dashed) and cotunneling theory (solid) predictions of the zero-bias, maximum Coulomb blockade conductance for a SET transistor as function of temperature. The simulation parameters are the same as those measured for the sample reported on.}
\end{figure}

Using the familiar Master Equation approach for the orthodox theory, we have calculated the curve shown in Fig. \ref{fig:Simulation} for a SET transistor with the parameters measured for the sample reported on here (JYU NM3). The figure shows the sequential tunneling and cotunneling conductances in zero bias and maximum Coulomb blockade, as function of temperature. It is seen that the conductances have the same order of magnitude, which means that sequential tunneling should be taken into account when interpreting the results.

\section{Experimental Setup\label{sec:setup}}

The samples are fabricated on 500 $\mu$m thick oxidized silicon. Gold pads and leads are made by a standard UV lithography technique. Before fabrication, the wafers are cleaned by ultrasound. A standard PMMA(-MAA) two-layer resist system is then applied and exposed by a pattern designed for 50 nm line with and 100 nm overlap. The line width becomes a little larger; 70-100 nm due to the exposure equipment and development. After exposure, the pattern is developed, and cleaned by reactive ion etching. Two layers of aluminium (40 and 30 nm thick) are evaporated at an angle, with an intermediate oxidation in 2 mbar O$_2$ for 3 minutes. After the last metal evaporation, the sample is oxidized again in 2 mbar O$_2$ for several minutes. Finally, the result is lifted off in acetone.

The samples were measured in the KelvinOx\texttrademark~dilution refrigerator at the Institute of Physics, DTU. The plastic mixing chamber has been replaced by a metal one, and the sample is mounted on a cold finger extending into a su\-per\-con\-duc\-ting magnet, which, in order to suppress the superconductivity of the aluminium, was set to one Tesla in all measurements presented here. The dilution refrigerator is set up in a shielded room, with the mechanical pumps outside the shielding. The temperature of the mixing chamber could be measured with calibrated germanium and ruthenium-oxide thermometers. The loaded base temperature is 50 mK. %Overvej? Ge vs. Ru

The samples were biased symmetrically using locally fabricated low-noise electronics. The electronics is mounted in an RF tight, compact metal box for shielding. For bias, an input amplifier adds two incoming signals; in this case a DC bias and a small low-frequency modulation provided by a lock-in amplifier. The added signal is then symmetrized by by an inverting and a non-inverting amplifier. Finally, the signal is voltage divided on each side to a suitable bias level $V_\mathrm{B}$, and fed to the sample through two large resistors $R_\mathrm{B}$. These were set to 10 M$\Omega$ in all measurements presented. A FET input, low-noise voltage preamplifier with a gain of 1000 was used to measure the resulting voltage $V_\mathrm{DC}$ across the sample, and the voltage signal is further amplified 10 times by an external PAR113 low-noise amplifier. The current through the sample is then calculated as $(V_\mathrm{B}-V_\mathrm{DC})/2R_\mathrm{B}$. A small-amplitude modulation, with frequency 2 Hz, was applied from the output of a SR850 lock-in amplifier on top of the DC bias to measure dynamic resistances. This results in a current modulation across the device of about 10 pA. After amplification in the FET preamplifier, the voltage response was fed back to the lock-in amplifier for measurement. With the measured zero-bias conductances, a 10 pA excitation corresponds in the worst case (2 $\mathrm{\mu S}$ differential conductance) to 50 aW heat input, and the voltage magnitude of the excitation is 5 $\mathrm{\mu V}$, corresponding to about 3\% of $e/C_\Sigma$. From the measured noise level at 2 Hz, combined with the equivalent noise bandwidth of the lock-in amplifier, we estimate the RMS noise to be at most 2.5\% of the signal responses obtained.

There are 20 DC connections from the room temperature electronics to the sample at the base temperature. Starting from the 300 K top flange, there are first 20 Thermocoax\textregistered~cables (length approx. 1.5 m) connect from room temperature to the 1.2 K level, then 20 superconducting wires in a ribbon to the mixing chamber (for thermal isolation), and finally 20 Thermocoax cables from the mixing chamber to the sample (length approx. 25 cm). The 20 Thermocoax cablesand the method of wiring minimize cross-talk and external noise input. The cables are thermally anchored at all temperature levels. The Thermocoax cables provide filtering of the room temperature thermal radiation;\cite{ZorinCoax} for $f$ = 10 GHz (corresponding to $h f/k_\mathrm{B}=0.5$ K) the attenuation of the cable is about 140 dB/m. The attenuation in dB increases as the square root of the frequency. Using a thermally conductive paste, the sample is mounted inside a thick-walled copper cavity and contacted electrically by ``buckling wires'' which tread on the sample. The 25 cm Thermocoax cables and the sample holder are enclosed in a copper shield at the mixing chamber temperature, which shields them from the surrounding 4.2 K radiation.

The individual shielding of the measurement leads also decreased the cross-capacitances considerably, which in turn minimizes cross-talk. The total capacitance to cryostat ground was measured to be 840 pF per lead at room temperature, which is in agreement with Zorin's result.\cite{ZorinCoax} It is worth noticing that the leakage resistance in these Thermocoax cables is quite low when the cryostat is at room temperature (down to tens of M$\Omega$), but increases to many G$\Omega$ when cooled down. We conjecture that this may be due to water vapor being absorbed in the MgO insulation powder at room temperature and pressure, but being cryopumped and/or frozen out when cooling the cryostat. This is important because the resistance of the SET transistor can be very high in the Coulomb blockade state. A matrix connector at room temperature and a ``switchboard'' (20 connectors which can be connected to any of the 20 slots connecting to the sample) at the mixing chamber further enables us to choose the very best cables for the critical connections. Cross-talk can only take place in the room temperature electronics, the matrix board, the superconducting ribbon, in the ``switchboard'' at the mixing chamber, and in the on-chip wiring.

To avoid microphonic pick-up from mechanic vibrations in the building (pumps etc.), the whole cryostat can be suspended using three inflated rubber tubes. Furthermore, lateral vibrations are damped with a fourth rubber tube concentric with the cryostat. This arrangement proved very efficient; components deriving from the asynchronous pumps (around 49 Hz) and other components at around 30 Hz vanish upon inflation of the rubber tubes.

The high capacitances in the filters and Thermocoax cables force us to use quite low modulation frequencies. For our symmetric biasing setup, the cut-off frequency is about $(G_\mathrm{D}+1/2R_\mathrm{B})/(2 \pi C_\ell)$, where $R_\mathrm{B}$ is the value of the the bias resistor, $G_\mathrm{D}$ is the differential conductance of the device, and $C_\ell$ is the total capacitance per lead (the 4-point biasing means that there are two leads contributing to the capacitance on each side). Because of the high internal cut-off frequency of the device, we find it reasonable to assume a simple resistive behaviour for the SET device at these frequencies. In the worst case, $G_\mathrm{D}$ is smaller than $1/2R_\mathrm{B}$, and in this case the cut-off will be at $1/ (4 \pi R_\mathrm{B} C_\ell)$. We have used $R_\mathrm{B}$ = 10 M$\Omega$ in all experiments reported here, giving a worst-case cut-off at 8 Hz. The modulation frequency was chosen to be 2 Hz, from which we have calculated the worst-case amplitude error due to the filters to be a few percent at the lowest conductances (about 2 $\mathrm{\mu S}$).

To apply microwaves to the device, there is a separate coaxial connection to the mixing chamber. At room temperature, a vacuum tight feedtrough connects an SMA connector to a standard 0.05'' 50 $\mathrm{\Omega}$ stainless steel coaxial cable inside the cryostat. A DC break is inserted before the vacuum feedthrough at room temperature, as well as before the thermal anchoring at 4.2 K. From this thermal anchor, the signal is fed through a short length of Thermocoax cable to another thermal anchor at the 1.2 K level. The Thermocoax cable serves as a cold attenuator (about 20 dB) for the microwave signal as well as unwanted signals such as room temperature radiation. The signal is then connected to the mixing chamber through a 20 m long superconducting 0.05'' 50 $\mathrm{\Omega}$ Nb coax cable, which is thermally anchored at the 0.6 K level. The superconductivity of the Nb provides a microwave connection while retaining thermal isolation of the mixing chamber. Before connecting to the mixing chamber, a third DC break is inserted in the line. All DC breaks are "inside-outside" breaks, meaning that both the inner and outer conductors are interrupted. The final microwave connection to the samples is provided by one of the 25 cm Thermocoax cables also used for the DC connections. This gives further attenuation of the signal.

As a consequence of the distributed attenuation and possible resonances in the microwave transmission line and wiring to the SET device, the actual power applied to the sample is unknown. It would me meaningless to make a throughput measurement at room temperature, since the damping characteristics are different at low temperature, and the coupling to the sample is unknown. Therefore, all power values are measured at the leveled output of the microwave synthesizer. The attenuation of the line at each frequency will be a fitting parameter in the data. For a given frequency, it is assumed that the actual power delivered to the SET device is a fixed fraction of the power from the high-frequency source.

Pressures and temperatures in the circulation system are continuously monitored by a computer, which also controls the biasing of the sample and reads the results measured by the voltage meters etc. through a GPIB bus. The GPIB bus is separated electrically from all measurement electronics by an optical link, which galvanically separates the computer from the measurement, and also enables the computer to be moved out of the shielded room if necessary.

\begin{figure}
\resizebox{\columnwidth}{!}{\includegraphics{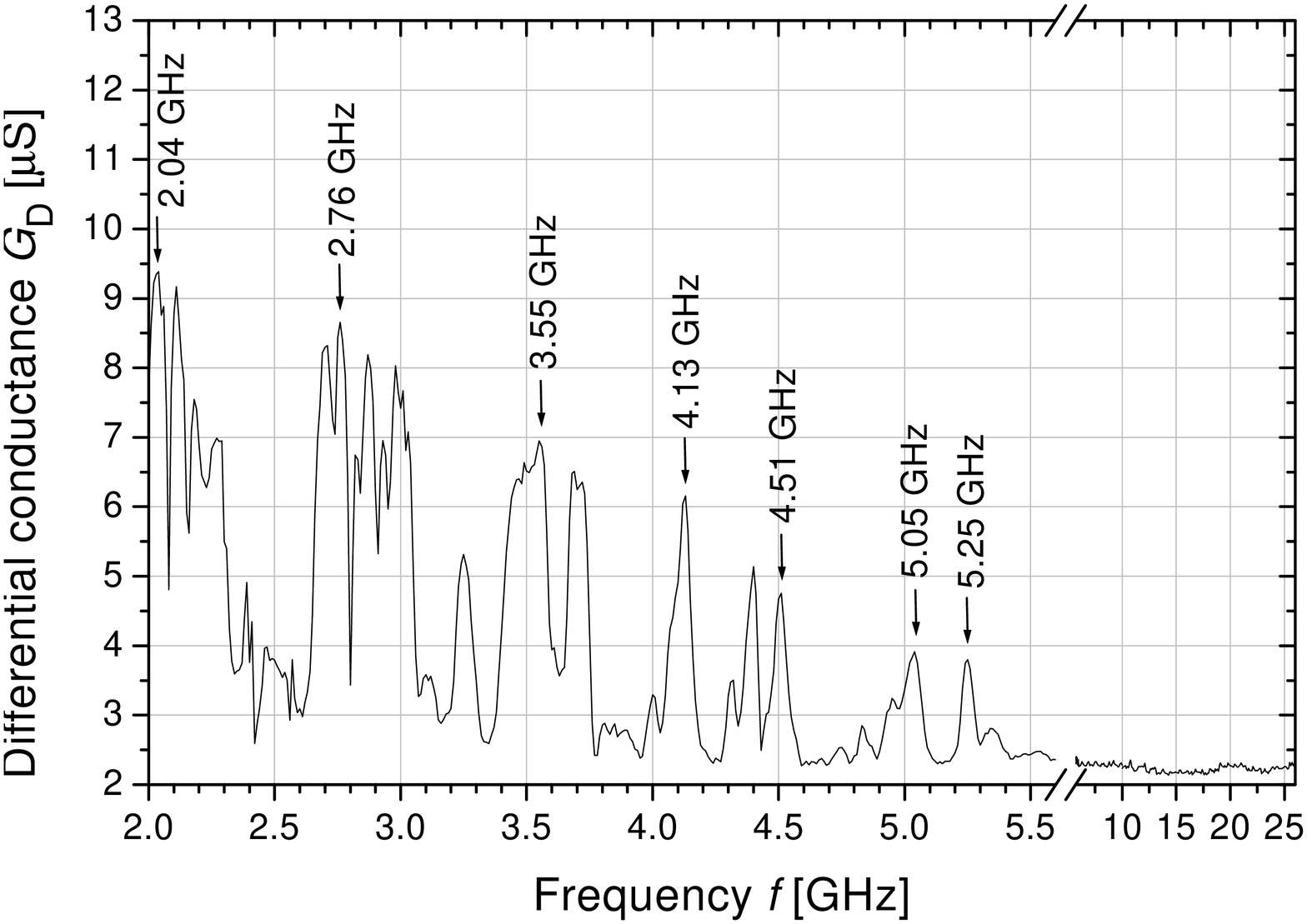}}%
\caption{\label{fig:Resonances} Resonances in the high-frequency line. The SET is used as a self-detector by biasing in zero bias, maximum blockade and applying a fixed $-10$ dBm signal from the microwave synthesizer. The differential conductance is then recorded as a function of the frequency. Notice the break in the frequency scale.}
\end{figure}

\section{Results\label{sec:results}}

\subsection{Device Characterization\label{sec:char}}

\begin{figure}
\resizebox{\columnwidth}{!}{\includegraphics{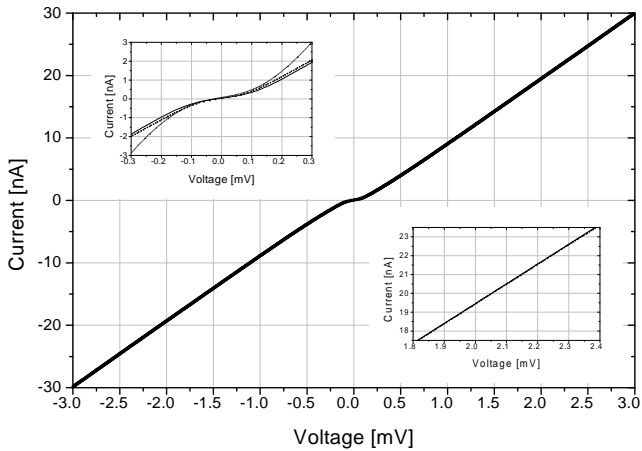}}%
\caption{\label{fig:I-V} Measured $I$-$V_\mathrm{DC}$ curve for sample JYU NM3 SET \#2 in maximum blockade, and the sequential tunneling prediction. The insets show enlargements of the blockade region (upper left) and the high-voltage region (lower right).}
\end{figure}

For the particular SET transistor reported on here, the DC $I$-$V_\mathrm{DC}$  curve (Fig. \ref{fig:I-V}) yield the device parameters \mbox{$R_\Sigma = R_1+R_2 = 95.0~\mathrm{k\Omega}$} and \mbox{$C_\Sigma = C_1+C_2 = 1030~\mathrm{aF}$}. From the $V_\mathrm{DC}$-$V_\mathrm{G}$ curves (not shown) we get \mbox{$C_\mathrm{G} = 0.94~\mathrm{aF}$}. The set of $V_\mathrm{DC}$-$V_\mathrm{G}$ curves at different current biases also provided evidence that the device is symmetric, which we will assume in the following. The particular sample was chosen from a batch for its charging energy, which also means a higher tunneling resistance (for fixed $RC$ product). Our simulations show that a high charging energy is more important than a low tunneling resistance in obtaining a high cotunneling/sequential tunneling ratio.

Since we want to first compare with the theory for zero current bias and maximum blockade, it is important to make sure that this is actually the case in the measurements. To do this, we use the following procedure: First, the maximum Coulomb blockade is found by sweeping across a peak in the $V_\mathrm{DC}$-$V_\mathrm{G}$ curve, and fitting to find the exact location of the peak. Then, the zero DC bias point was found in a similar manner by recording and $I$-$V_\mathrm{DC}$ curve around zero DC bias and fitting the dynamic resistance to find the peak. This procedure was carried out before each measurement.

The microwave connection obviously exhibits resonances (see section \ref{sec:setup}), at which the signal received by the SET is higher. It is advantageous for us to utilise these resonance frequencies for the microwave bias, and to determine them we used the SET device as a self-detector. It was assumed that the frequency dependence of the device response was small in the relatively narrow frequency range probed. The resonances are determined by measuring the dynamic resistance at zero DC bias and maximum blockade (using the method described above) in small steps. The method is based on the assumption of a response both from the sequential tunneling and cotunneling contributions, as neither of these are linear around zero bias. The result is shown in Fig. \ref{fig:Resonances}, where some clear peaks in the dynamic resistance are seen at various frequencies. The results shown in this paper are measured at the frequencies marked by arrows.

\subsection{Temperature of the measurements}\label{sec:Temperature}

\begin{figure}
\resizebox{\columnwidth}{!}{\includegraphics{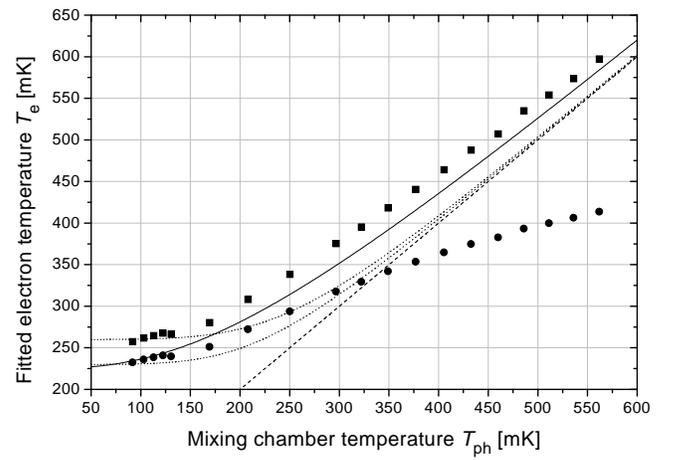}}%
\caption{\label{fig:Temperatures} Temperature estimates from the zero-bias conductance. The fit to the sequential tunneling only (squares) provides an upper bound, while the fit to sequential tunneling plus cotunneling (points) is a lower bound. The electron temperature should approach the mixing chamber temperature (dashed line) at high temperatures. Also shown are the forms $T_\mathrm{e} = \sqrt[3]{\smash[b]{T_0^3+T_\mathrm{ph}^3}}$ (solid line;  $T_0$ = 227 mK) and $T_\mathrm{e} = \sqrt[5]{\smash[b]{T_0^5+T_\mathrm{ph}^5}}$ (dotted lines; $T_0$ = 260 mK for the upper curve, $T_0$ = 235 mK for the lower curve).}
\end{figure}

An important question relating to any measurement at low temperature is: What is the correct temperature? First of all, the sample may be far away from the cooling source; in our case the mixing chamber of the dilution refrigerator. This means that any heat delivered to the sample by Joule heating or radiation may give a higher temperature at the sample than at the mixing chamber. Second, there is the question of whether the electron system is in thermal equilibrium with the phonon system.\cite{Wellstood} This suggests that it is not correct to rely on the temperature measured by the thermometers at the mixing chamber, even when these are well-calibrated.

To get an estimate of the real temperature temperature of the electron system in the SET transistor, we use the results of the differential conductance measurements at maximum blockade and zero bias. Using both the cotunneling theory (Eq. \ref{eq:ACcotun3rdOrder}) and the orthodox theory, the temperature has been fitted to make the differential conductance $G_\mathrm{D}$ agree. The junction parameters were fixed at the values determined in the previous section, leaving $T_\mathrm{e}$ as the only free parameter. The results of this temperature fitting is shown in Fig. \ref{fig:Temperatures} as function of the mixing chamber temperature $T_\mathrm{ph}$ (which we, following Ref. \onlinecite{Wellstood}, assume equal to the phonon temperature, hence the symbol). The temperature has been fitted both to the sequential tunneling conductance alone (squares), which will give an upper bound on the temperature, and to the tunneling conductance with cotunneling included (circles), which will give a lower bound under the assumption that other effects can be neglected. Thus in the last case we have implicitly assumed that we can use the theory for cotunneling by Averin \& Nazarov at low temperatures, which has been confirmed by several researchers.\cite{Geerlings,Eiles} Also shown is the ``strong coupling'' curve $T_\mathrm{e}=T_\mathrm{ph}$ (dashed line), which can be considered a ``hard'' lower bound.

It is seen that the fitted electron temperatures indeed saturate at low mixing chamber temperatures, as suggested in the work by Wellstood et {\em al.}\cite{Wellstood} However, attempts to make the results fit to the $T_\mathrm{e} = \sqrt[5]{\smash[b]{T_0^5+T_\mathrm{ph}^5}}$ form, which applies for uniform heating in a thin film, seem to fail for our SET structure ($T_0$ is the temperature at which $T_\mathrm{e}$ saturates for $T_\mathrm{ph} \rightarrow 0$). A much more reasonable fit is provided by the form $T_\mathrm{e} = \sqrt[2.5]{\smash[b]{T_0^{2.5}+T_\mathrm{ph}^{2.5}}}$ (solid line). This form respects the lower bound as well as the limit at $T_\mathrm{ph} \rightarrow 0$. Such a dependence is not entirely unreasonable, considering that the heating is localized in the SET transistor, and that heat transfer is provided only by the leads, which should probably be considered between one- and two-dimensional. Indeed the experiments of Wellstood et {\it al.} on local heating of a SQUID with cooling fins gave a $T_\mathrm{e}^{2.7}$ form.

We will adopt the $T_\mathrm{e} = \sqrt[2.5]{\smash[b]{T_0^{2.5}+T_\mathrm{ph}^{2.5}}}$ form as our best estimate of the temperature in the following. We make no claims that this is the accurate temperature, only that it seems to be a reasonable approximation which makes a sensible transition from the low-temperature to the high-temperature regime.

\subsection{$V_\mathrm{RF}^2$ dependence\label{sec:Vac}}

In this section, we present measurements at maximum blockade, which means that $\Delta^\pm$ reduce to the charging energy $E_\mathrm{C}$. Also, since there is no asymmetry in the device at zero bias and maximum blockade (no tunneling direction is favored), there should be no frequency dependence in the differential conductance, which is also predicted by theory. The absence of frequency dependence enables us to find the relative damping of the line at the individual frequencies. Without frequency dependence, the zero bias conductance $G_\mathrm{D}$ as function of the microwave amplitude $V_\mathrm{RF}$ should be the same at all frequencies. This means that the relative damping can be found by minimizing the mean square difference between the $G_\mathrm{D}$-$V_\mathrm{RF}$ curves at the individual frequencies, using the damping as free parameter. This procedure provided us with the relative damping values 0.0, 3.8, 9.1, 11.5, 14.8 17.3, and 17.6 dB for the frequencies marked in Fig. \ref{fig:Resonances}. As expected from Fig. \ref{fig:Resonances}, the damping increases with frequency. However, it should be noted that the optimal values vary slightly with temperature, suggesting that there might be a co-dependence on temperature and frequency.

\begin{figure}
\resizebox{\columnwidth}{!}{\includegraphics{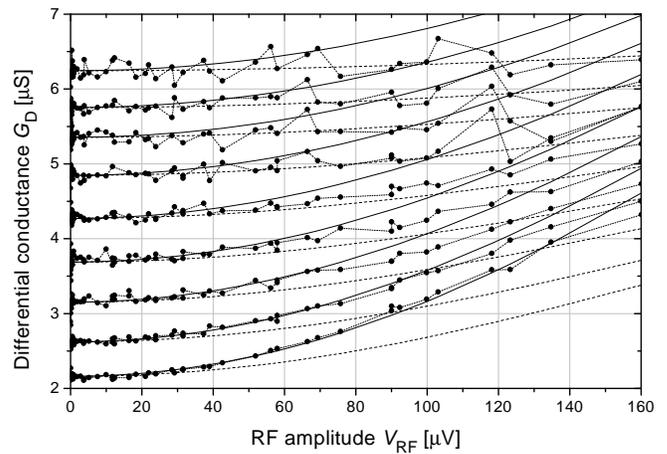}}%
\caption{\label{fig:Gd-Vac37.9dB} Differential conductance as function of the microwave amplitude at 37.9 dB additional damping. The measured values (points) are connected to guide the eye. The lines are the measured zero-bias value plus the predicted excess conductance with cotunneling (solid) and without cotunneling (dashed). The predictions are at 232, 255, 275, 301, 336, 379, 428, 478, and 549 mK, respectively (bottom to top).}
\end{figure}

The additional damping (i.e. the damping that should be added to the relative damping to get the absolute damping) should be the same at all frequencies and temperatures. The absolute damping is unfortunately unknown. All we can do is to find a damping value that is consistent with theory. In Fig. \ref{fig:Gd-Vac37.9dB}, the resulting theoretical prediction of the $G_\mathrm{D}$-$V_\mathrm{RF}$ curves at different temperatures are shown along with the measured values, assuming an additional damping of 37.9 dB. It is seen that the curves from different frequencies have indeed collapsed into one, especially at the lowest temperature. At higher temperatures, where the change in tunneling is more modest, the picture is not as perfect. Also shown is the predicted excess conductance with and without cotunneling at the same temperature added to the zero-power conductance. It is seen that the measured conductance fits the predicted one quite well, especially at low temperatures.

\begin{figure}
\resizebox{\columnwidth}{!}{\includegraphics{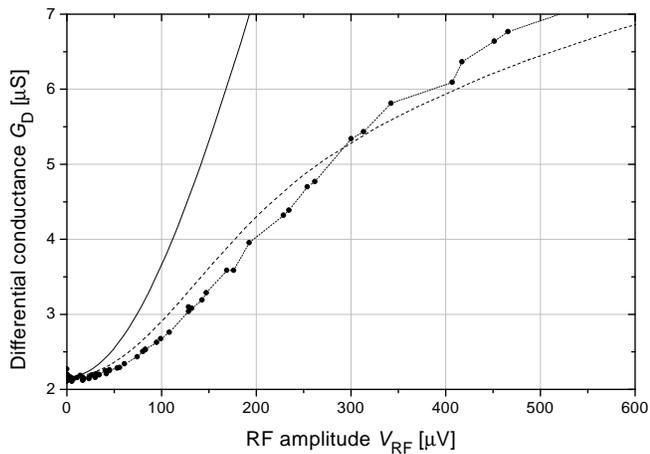}}%
\caption{\label{fig:Gd-Vac36.2dB} Differential conductance as function of microwave amplitude at 36.2 dB additional damping. The data (points) are the same as the bottom curves in Fig. \ref{fig:Gd-Vac37.9dB}, except that a lower damping in the microwave line has been assumed. The lines are the measured zero-bias value plus the predicted excess conductance with cotunneling (solid) and without cotunneling (dashed), for a temperature of 257 mK. This is the electron temperature fitted from the maximum blockade conductance (compare the squares in Fig. \ref{fig:Temperatures}). Note that the horizontal scale is different from that in Fig. \ref{fig:Gd-Vac37.9dB}.}
\end{figure}

The results just presented could lead to the hasty conclusion that the theory for coherent photon assisted cotunneling is correct. However, there are several points which cause concern. First, the additional damping required to make the curves fit changes with temperature, which it should not, unless the change from about 200 mK to about 500 mK really increases the damping (which seems very improbable). There could be a range of other explanations for this result. For example, the change could be an artefact of the fitting procedure: If the excess cotunneling conductance becomes smaller than predicted by theory at higher temperatures (which seems reasonable), this will show up in the fit as a higher damping. A related problem is that the quality of the fit should be the same at all temperatures if the temperature used for the theoretical prediction is correct (i.e. equal to the actual electron temperature). This is because the temperature does not (in the theory) contribute to the {\em excess} cotunneling conductance; it only gives a constant contribution (compare Eq. \ref{eq:ACcotun3rdOrder}). However, the fit is best at low temperatures.

One `sanity check' that should be performed is whether the orthodox theory alone can explain the results. Using the value 36.2 dB for the additional damping gives the result in Fig. \ref{fig:Gd-Vac36.2dB}. It is seen that the results cross the prediction for the orthodox tunneling at this damping value. At higher damping, the conductance would be too high at high amplitudes, while a lower damping the conductance would be too low at low amplitudes. Thus the results can not be explained by the orthodox theory, even if one assumes a higher electron temperature and another damping. The electron temperature would also have to be substantially higher than at the mixing chamber at all temperatures, as seen in Fig. \ref{fig:Temperatures}. Note that although it would seem that we are applying a double standard here, since the horizontal scale is different in Fig. \ref{fig:Gd-Vac37.9dB} and \ref{fig:Gd-Vac36.2dB}. However, the orthodox theory should be valid for all voltages and thus amplitudes, while the cotunneling theory only claims validity for $e V_\mathrm{RF} \ll \Delta^\pm$. 

\subsection{The $\Delta^\pm$ dependence}

	\begin{figure}
	\resizebox{\columnwidth}{!}{\includegraphics{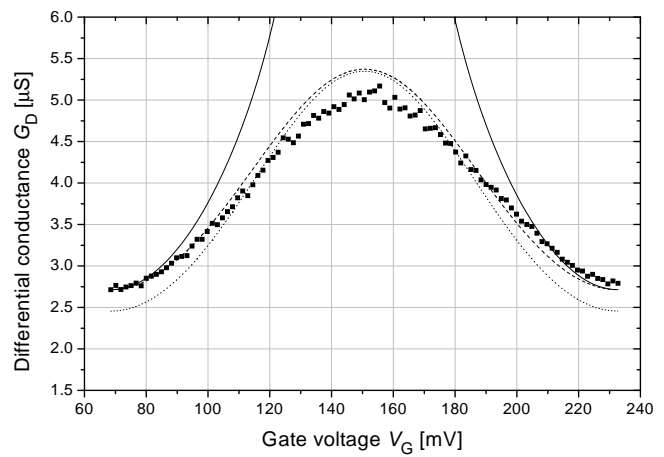}}
	\caption{\label{fig:Delta}The $\Delta^\pm$ dependence (points) and the predicted excess conductance added to the measured zero-bias conductance, with and without cotunneling (solid and dashed lines, respectively). The prediction with cotunneling is made at 272 mK, which is the assumed electron temperature ($T^{2.5}$ form; line in Fig. \ref{fig:Temperatures}). The sequential tunneling prediction is made at 286 mK; this is the temperature that makes the the maximum-blockade point fit (like the squares in Fig. \ref{fig:Temperatures}). Also shown is the sequential tunneling part of the prediction with cotunneling (dotted line).}
	\end{figure}

The dependence of the cotunneling on $\Delta^\pm$ has been investigated by measuring the dynamic conductance at zero bias as function of $V_\mathrm{G}$. Zero bias was found as described previously, and the device was biased within one $e$-period of the gate bias. The result is shown in Fig. \ref{fig:Delta}, along with the predictions of the excess conductance with and without cotunneling added to the maximum-blockade result. For the prediction including cotunneling, the assumed electron temperature is used (solid line in Fig. \ref{fig:Temperatures}); for the sequential tunneling the temperature fitting the maximum-blockade point (squares in Fig. \ref{fig:Temperatures}). As expected, the cotunneling prediction fits reasonably well near maximum blockade, but diverges quickly when approaching minimum blockade, where $\Delta^\pm \rightarrow 0$. Also shown is the sequential tunneling contribution to the current; it is seen that if the temperature used is correct, the sequential tunneling more or less `takes over' as  $\Delta^\pm$ approaches zero. The sequential tunneling alone seems to fit surprisingly well. One explanation for the good fit is that as $\Delta^\pm$ approaches zero, the threshold voltage becomes smaller,\cite{Eiles} and thus the orthodox theory will account for most of the current. %overvej dette, omskriv evt.

\subsection{Photon assisted cotunneling away from blockade}

The theory derived by Flensberg \cite{Flensberg} applies for $e V_\mathrm{DC},\:k_\mathrm{B} T_\mathrm{e},\:e V_\mathrm{RF},\:hf\ll \Delta^\pm$. Thus, when one of $\Delta^\pm$ becomes smaller, the voltage/temperature/amplitude interval where this assumption is valid becomes smaller. Measurements similar to those presented in the previous sections have been performed also at $C_\mathrm{G} V_\mathrm{G}/e = 0.25,0.50,0.75$, i.e. $\Delta^\pm/E_\mathrm{C} = 0.5,0.0,0.5$. Thus the bounds of validity for $V_\mathrm{DC}$, $T_\mathrm{e}$ and $V_\mathrm{RF}$ are diminished at $C_\mathrm{G} V_\mathrm{G}/e = 0.25,0.75$ (slopes in the $V_\mathrm{DC}$-$V_\mathrm{G}$ curve), while the condition is impossible to satisfy at $C_\mathrm{G} V_\mathrm{G}/e = 0.50$ (minimum blockade), where $\Delta^\pm = 0$. Another way of stating this is that Eq. \ref{eq:ACcotun3rdOrder} diverges at $\Delta^\pm \rightarrow 0$.

\begin{figure}
	\begin{tabular}{c}
	\resizebox{\columnwidth}{!}{\includegraphics{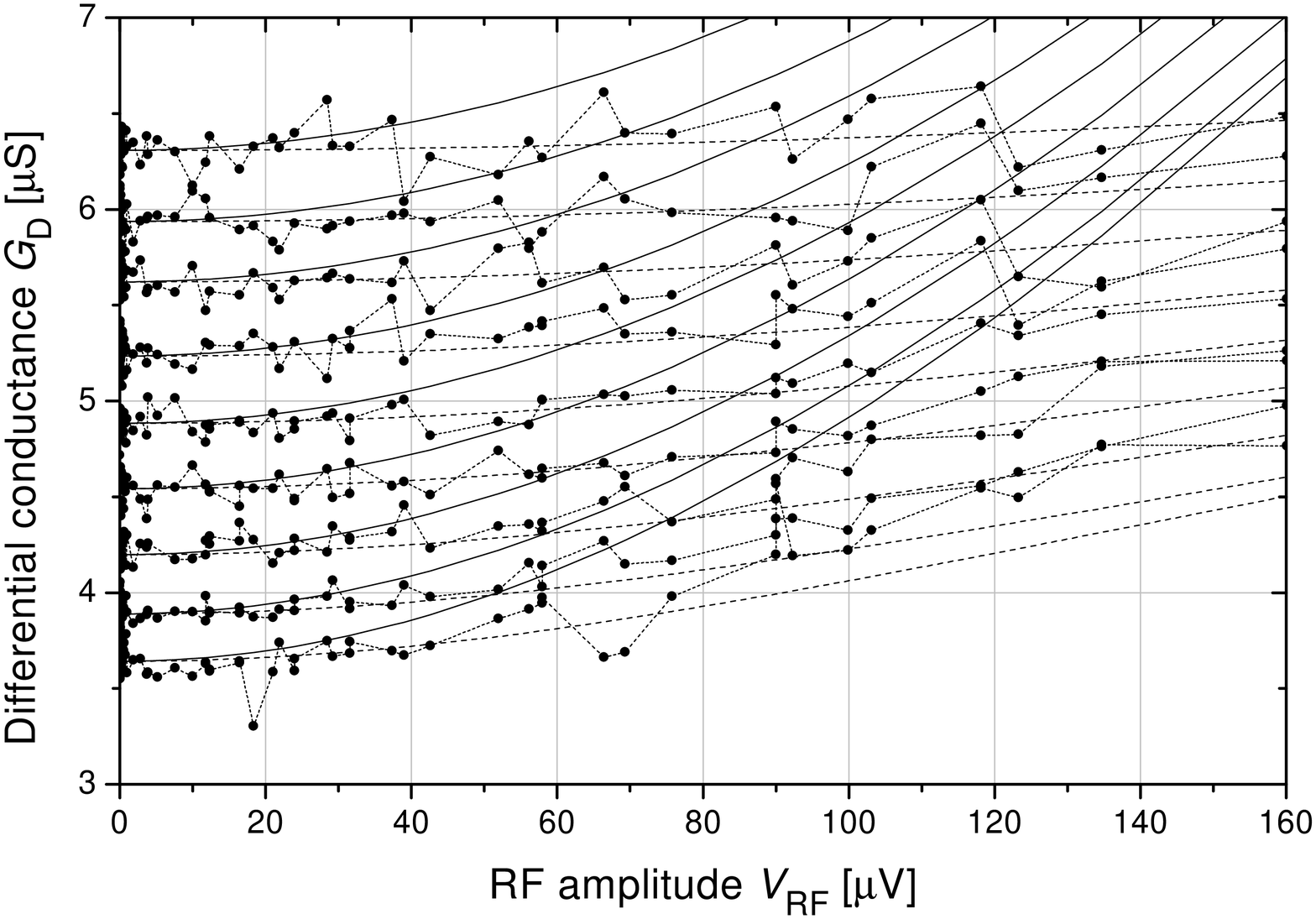}} \\
	\resizebox{\columnwidth}{!}{\includegraphics{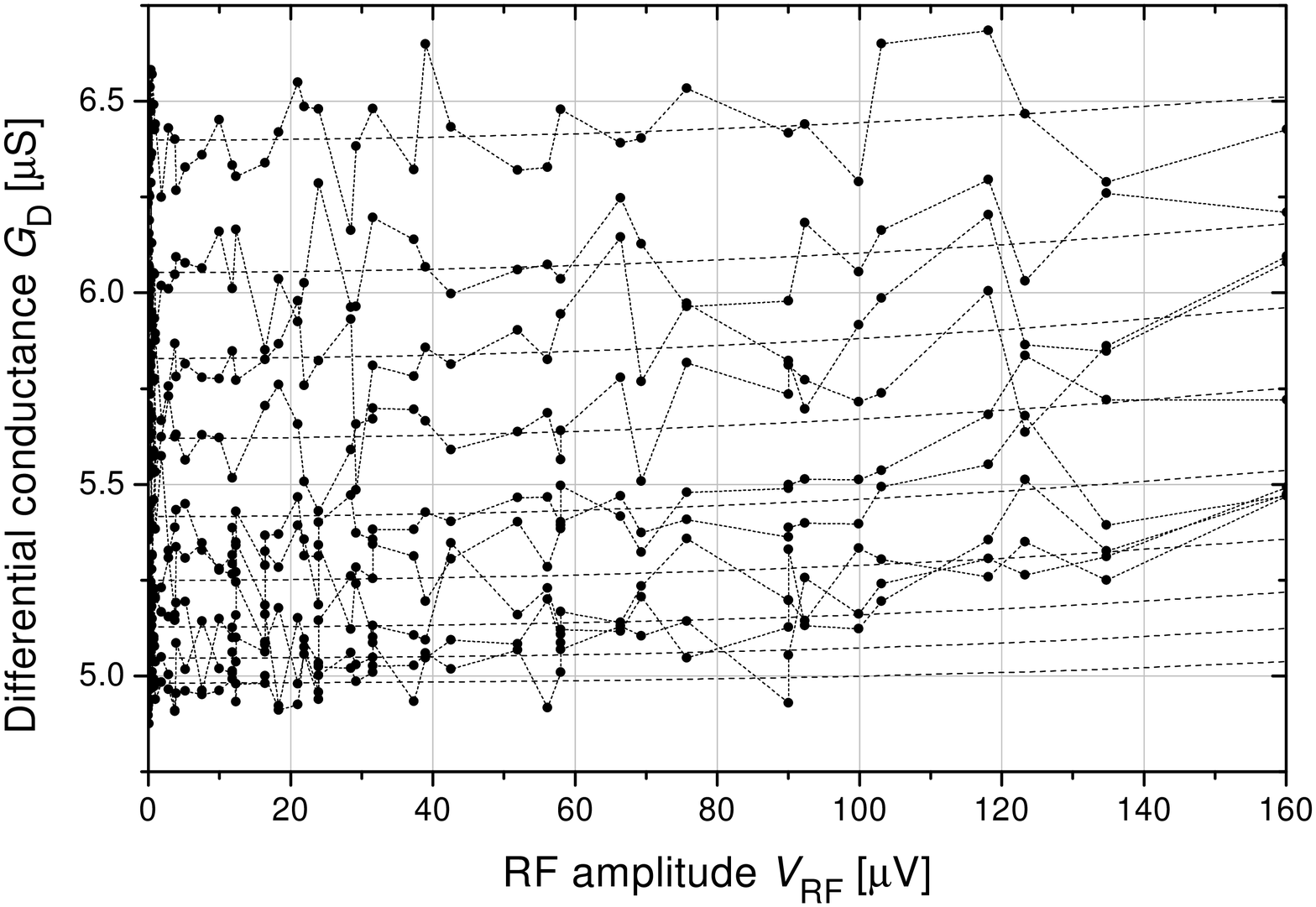}} \\
	\resizebox{\columnwidth}{!}{\includegraphics{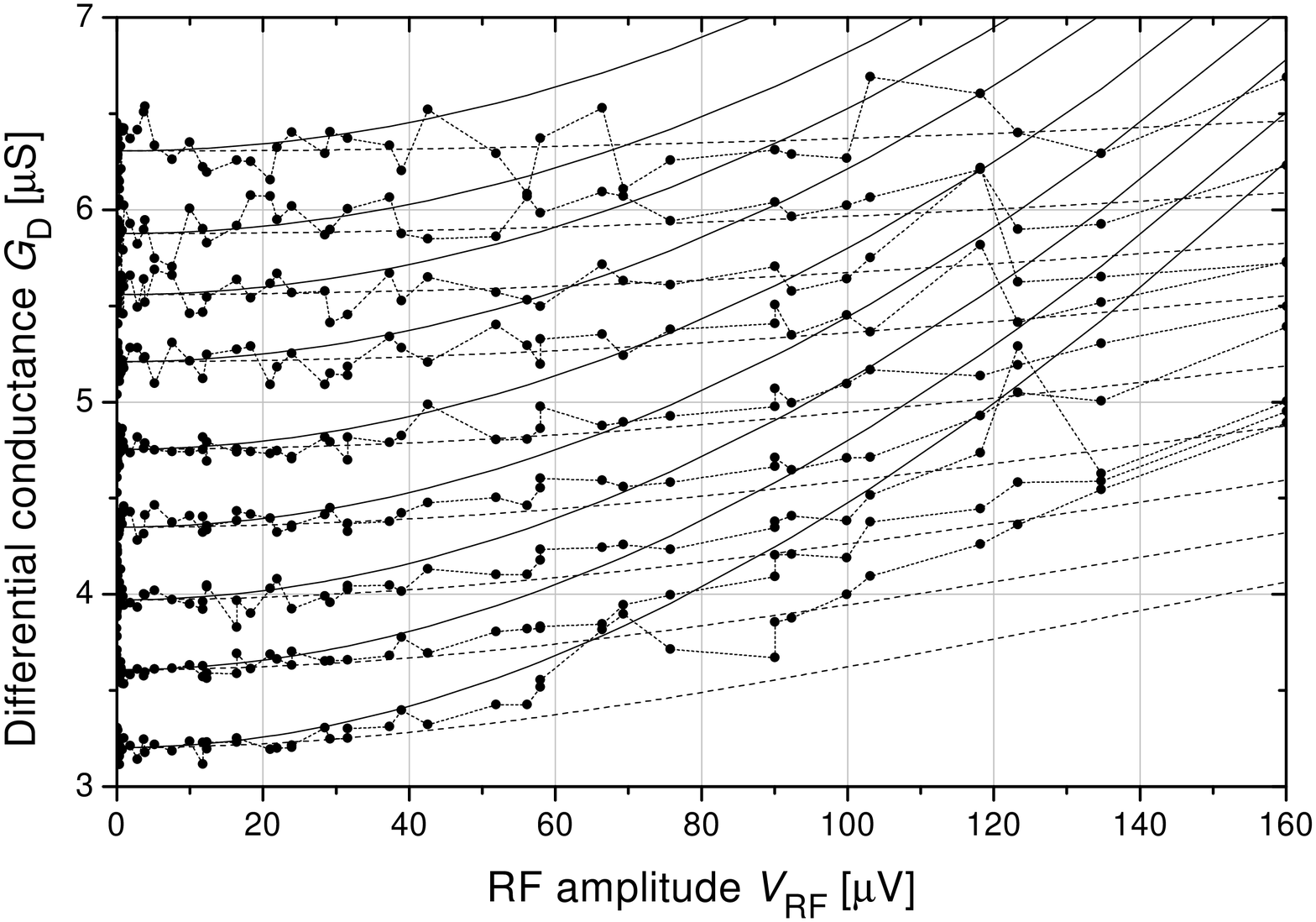}} \\
	\end{tabular}
	\caption{\label{fig:NonBlockade} Microwave assisted cotunneling away from blockade, using the results at the seven frequencies collected by applying their relative damping. The curves are for the SET transistor biased at the negative slope of the $V_\mathrm{DC}$-$V_\mathrm{G}$ curve (top), minimum blockade (middle) and positive slope (bottom). Note the different vertical scale in the middle graph. The temperatures, line types etc. are the same as in Fig. \ref{fig:Gd-Vac37.9dB}.}
	\end{figure}

The results are presented in Fig. \ref{fig:NonBlockade}. It is seen that on the slopes, the observed conductance falls between the predictions with and without cotunneling, which is expected since the cotunneling should be overestimated at these values of $\Delta^\pm$. At minimum blockade the conductance approaches the prediction from the orthodox theory alone, which is reasonable for a SET transistor in the minimum blockade, where the sequential tunneling dominates (this is similar to approaching the charging voltage at maximum blockade).

\section{Discussion\label{sec:discussion}}

The interpretation of the results involves several explicit and implicit assumptions. To determine the lowest temperatures, it is assumed that both the orthodox theory for sequential tunneling and the original cotunneling theory by Averin \& Nazarov \cite{AverinNazarov} were correct within the given bounds of validity. The orthodox theory has been extensively confirmed since the first theories were published, and also the original cotunneling has been verified.\cite{Geerlings,Eiles} It is further assumed that the coupling from the electron system to the mixing chamber through the phonon system was strong enough at high temperatures to assume a convergence towards the mixing chamber temperature here.

%Gentagelse?
The largest problem in the interpretation of the results is that both the electron temperature of the device and the damping in the microwave line were not known exactly. Since it is certainly insufficient to use the mxing chamber temperature, which is not in equilibrium with the electron temperature, it is necessary to rely on existing theories and some assumptions to get an estimate of the electron temperature. Similarly, it is necessary to rely on the absence of frequency dependence in the device to determine the relative damping values at the different frequencies. The additional damping is then an adjustable parameter. Here it is implicitly assumed that the damping does not depend on amplitude. Ideally, it should also have been checked thoroughly that the differential conductance measurement was independent of the modulation amplitude. However, this was only done sporadically, i.e. by halving and doubling the amplitude and checking that the response followed.

Another problem is that the temperatures did not range to below about $T_\mathrm{C}/4$ (another way of saying this is that the tunnel capacitances were somewhat too high). Thus it is difficult to say with conviction that the condition $k_\mathrm{B} T_\mathrm{e} \ll \Delta^\pm$ is satisfied, even at the lowest temperatures.

Among the other assumptions are that the the SET transistor exhibited a resistive behaviour at low frequencies (should be reasonable since $R_\Sigma C_\Sigma$ corresponds to about 1.7 GHz), that the damping in the microwave line is linear, that the leads are equal (small correction), and that other effects, such as noise and the environment, can be neglected. % Reference på vej fra DTV

Under the given assumptions, the damping could be chosen such that results are consistent with the theory at low temperatures. Furthermore, it is shown that sequential tunneling alone could not explain the results. Thus, the results speak in favor of the cotunneling theory. At least we can say that more than the orthodox theory is needed to explain the results, and that cotunneling is a resonable explanation. However, due to the dicrepancies observed, it is clear that a more complete theory is needed.

There is still the unsettling matter of the required additional damping given by the fitting procedure is changing with temperature. A number of explanations can be offered for this. It could be the coupling that changes, but it is hard to see how such a small change in temperature (from 200 mK to 500 mK) would cause a change of 10 dB in the damping. A more reasonable explanation could be that the enhancement of cotunneling by a coherent source must be temperature dependent (i.e. if the real temperature contribution becomes smaller than predicted by the theory, this will in the fitting be compensated by increasing the damping, making the microwave contribution smaller instead), which would require the assumed electron temperatures to be approximately correct. For example, we may conjecture that at higher temperatures, the limit of validity is not far away; i.e. that we should require something like $(2\pi k_\mathrm{B} T_\mathrm{e})^2 + (eV_\mathrm{DC})^2 + \frac{3}{2}(e V_\mathrm{RF})^2 \ll (\Delta^\pm)^2$ in Eq. \ref{eq:ACcotun3rdOrder}. Another possibility is that the temperatures used are too low, with the error increasing with temperature. This would make the orthodox prediction too high, which would then be compensated in the fit by a higher damping. However, the orthodox theory prediction approaches the conjectured electron temperatures at high temperatures, making the error margin upwards smaller. 

Also the relative damping varied with temperature, hinting that also the frequency might have a different effect on the conductance at different temperatures, again contrary to the theoretical prediction. Indeed the charging frequency of the device is $f_\mathrm{C}=18.8$ GHz, which means that the frequencies used range from $0.11f_\mathrm{C}$ to $0.28f_\mathrm{C}$. Similar arguments as above could be applied to this matter.

\section{Conclusion and Outlook\label{sec:conclusion}}

The enhancement of cotunneling by a coherent microwave source was investigated experimentally and compared with theory. The results on cotunneling seem to confirm the theory for adiabatic enhancement of cotunneling by a coherent microwave source. However, the presence of assumptions and adjustable parameters should inspire modesty in the strength of the conclusions drawn. Furthermore, the results seem to suggest that the effect of microwave amplitude and temperature are not independent, contrary to what is suggested by theory. Thus a more complete theory is needed. Experimentalists would very much appreciate a theory which gives a method of actually calculating the cotunneling current at all voltages, temperatures etc. (e.g. something like the Master Equation approach). In any case, the presence of microwave induced cotunneling enhances the importance of taking cotunneling into account in the calculation of SET devices.

It is vital in future measurements to have samples with a high charging energy to make the charging temperature higher. Also one could attempt to lower the electron temperature, e.g. by using cooling fins on the sample. Cold resistors and cross-correlation voltage measurements should be used to improve the noise. A new voltage bias electronics, which will reduce the noise and eliminate the problem with resistive biasing of a high-impedance sample, is under construction. The measurements should focus on verifying the photon-assisted current in Eq. \ref{eq:ACcotun4thOrder}, e.g. by biasing in zero voltage and observing the current as function of amplitude and frequency. A careful determination of the device parameters and damping of the microwave line is essential, as there will be adiabatic contributions from both sequential tunneling and cotunneling. It should futher be checked that the differential conductance result is idependent of the amplitude of the lock-in modulation. Also the measurements should be performed on many more samples to rule out sample-specific effects.

%\appendix
%\section{}

\begin{acknowledgments}
The authors would like to thank the SET groups at PTB for many fruitful discussions. Mikkel Ejrn\ae s is gratefully acknowledged for his work on improving the cryostat and the computer measurements. Karsten Flensberg is thanked for clarifying discussions and thoughtful comments. The project was supported in parts by the Danish Natural Science Foundation, the Hartmann Foundation, and EU (contract IST-1999-10673). MS acknowledges a grant from Tekniikan Edist\"amiss\"a\"ati\"o (TES).
\end{acknowledgments}

\bibliography{PRB}

\begin{thebibliography}{18}
\expandafter\ifx\csname natexlab\endcsname\relax\def\natexlab#1{#1}\fi
\expandafter\ifx\csname bibnamefont\endcsname\relax
  \def\bibnamefont#1{#1}\fi
\expandafter\ifx\csname bibfnamefont\endcsname\relax
  \def\bibfnamefont#1{#1}\fi
\expandafter\ifx\csname citenamefont\endcsname\relax
  \def\citenamefont#1{#1}\fi
\expandafter\ifx\csname url\endcsname\relax
  \def\url#1{\texttt{#1}}\fi
\expandafter\ifx\csname urlprefix\endcsname\relax\def\urlprefix{URL }\fi
\providecommand{\bibinfo}[2]{#2}
\providecommand{\eprint}[2][]{\url{#2}}

\bibitem[{\citenamefont{Kautz et~al.}(1999)\citenamefont{Kautz, Keller, and
  Martinis}}]{Kautz}
\bibinfo{author}{\bibfnamefont{R.~L.} \bibnamefont{Kautz}},
  \bibinfo{author}{\bibfnamefont{M.~W.} \bibnamefont{Keller}},
  \bibnamefont{and} \bibinfo{author}{\bibfnamefont{J.~M.}
  \bibnamefont{Martinis}}, \bibinfo{journal}{Phys. Rev. B}
  \textbf{\bibinfo{volume}{60}}, \bibinfo{pages}{8199} (\bibinfo{year}{1999}).

\bibitem[{\citenamefont{Keller et~al.}(1999)\citenamefont{Keller, Eichenberger,
  Martinis, and Zimmerman}}]{Keller}
\bibinfo{author}{\bibfnamefont{M.~W.} \bibnamefont{Keller}},
  \bibinfo{author}{\bibfnamefont{A.~L.} \bibnamefont{Eichenberger}},
  \bibinfo{author}{\bibfnamefont{J.~M.} \bibnamefont{Martinis}},
  \bibnamefont{and} \bibinfo{author}{\bibfnamefont{N.~M.}
  \bibnamefont{Zimmerman}}, \bibinfo{journal}{Science}
  \textbf{\bibinfo{volume}{285}}, \bibinfo{pages}{1706} (\bibinfo{year}{1999}).

\bibitem[{\citenamefont{Aassime et~al.}(2001)\citenamefont{Aassime, Gunnarsson,
  Bladh, Delsing, and Schoelkopf}}]{Aassime}
\bibinfo{author}{\bibfnamefont{A.}~\bibnamefont{Aassime}},
  \bibinfo{author}{\bibfnamefont{D.}~\bibnamefont{Gunnarsson}},
  \bibinfo{author}{\bibfnamefont{K.}~\bibnamefont{Bladh}},
  \bibinfo{author}{\bibfnamefont{P.}~\bibnamefont{Delsing}}, \bibnamefont{and}
  \bibinfo{author}{\bibfnamefont{R.~J.} \bibnamefont{Schoelkopf}},
  \bibinfo{journal}{Applied Physics Letters} \textbf{\bibinfo{volume}{79}},
  \bibinfo{pages}{4031} (\bibinfo{year}{2001}).

\bibitem[{\citenamefont{Vion et~al.}(2002)\citenamefont{Vion, Aassime, Cottet,
  Joyez, Pothier, Urbina, Esteve, and Devoret}}]{Vion}
\bibinfo{author}{\bibfnamefont{D.}~\bibnamefont{Vion}},
  \bibinfo{author}{\bibfnamefont{A.}~\bibnamefont{Aassime}},
  \bibinfo{author}{\bibfnamefont{A.}~\bibnamefont{Cottet}},
  \bibinfo{author}{\bibfnamefont{P.}~\bibnamefont{Joyez}},
  \bibinfo{author}{\bibfnamefont{H.}~\bibnamefont{Pothier}},
  \bibinfo{author}{\bibfnamefont{C.}~\bibnamefont{Urbina}},
  \bibinfo{author}{\bibfnamefont{D.}~\bibnamefont{Esteve}}, \bibnamefont{and}
  \bibinfo{author}{\bibfnamefont{M.~H.} \bibnamefont{Devoret}},
  \bibinfo{journal}{Science} \textbf{\bibinfo{volume}{296}},
  \bibinfo{pages}{886} (\bibinfo{year}{2002}).

\bibitem[{\citenamefont{Kauppinen et~al.}(1998)\citenamefont{Kauppinen, Loberg,
  Manninen, and Pekola}}]{Kauppinen}
\bibinfo{author}{\bibfnamefont{J.~P.} \bibnamefont{Kauppinen}},
  \bibinfo{author}{\bibfnamefont{K.~T.} \bibnamefont{Loberg}},
  \bibinfo{author}{\bibfnamefont{A.~J.} \bibnamefont{Manninen}},
  \bibnamefont{and} \bibinfo{author}{\bibfnamefont{J.~P.}
  \bibnamefont{Pekola}}, \bibinfo{journal}{Rev. Sci. Instrum.}
  \textbf{\bibinfo{volume}{69}}, \bibinfo{pages}{4166} (\bibinfo{year}{1998}).

\bibitem[{\citenamefont{Bergsten et~al.}(2001)\citenamefont{Bergsten, Claeson,
  and Delsing}}]{Bergsten}
\bibinfo{author}{\bibfnamefont{T.}~\bibnamefont{Bergsten}},
  \bibinfo{author}{\bibfnamefont{T.}~\bibnamefont{Claeson}}, \bibnamefont{and}
  \bibinfo{author}{\bibfnamefont{P.}~\bibnamefont{Delsing}},
  \bibinfo{journal}{Appl. Phys. Lett.} \textbf{\bibinfo{volume}{78}},
  \bibinfo{pages}{1264} (\bibinfo{year}{2001}).

\bibitem[{\citenamefont{Averin and Likharev}(1991)}]{AverinLikharev}
\bibinfo{author}{\bibfnamefont{D.~V.} \bibnamefont{Averin}} \bibnamefont{and}
  \bibinfo{author}{\bibfnamefont{K.~K.} \bibnamefont{Likharev}}, in
  \emph{\bibinfo{booktitle}{Mesoscopic Phenomena in Solids}}, edited by
  \bibinfo{editor}{\bibfnamefont{B.~L.} \bibnamefont{Altshuler}},
  \bibinfo{editor}{\bibfnamefont{P.~A.} \bibnamefont{Lee}}, \bibnamefont{and}
  \bibinfo{editor}{\bibfnamefont{R.~A.} \bibnamefont{Webb}}
  (\bibinfo{publisher}{Elsevier}, \bibinfo{year}{1991}),
  vol.~\bibinfo{volume}{30} of \emph{\bibinfo{series}{Modern Problems in
  Condensed Matter Physics}}, chap.~\bibinfo{chapter}{6}, pp.
  \bibinfo{pages}{173--271}.

\bibitem[{\citenamefont{Ingold and Nazarov}(1992)}]{IngoldNazarov}
\bibinfo{author}{\bibfnamefont{G.-L.} \bibnamefont{Ingold}} \bibnamefont{and}
  \bibinfo{author}{\bibfnamefont{Y.~V.} \bibnamefont{Nazarov}}, in
  \emph{\bibinfo{booktitle}{Single Charge Tunneling}}, edited by
  \bibinfo{editor}{\bibfnamefont{H.}~\bibnamefont{Grabert}} \bibnamefont{and}
  \bibinfo{editor}{\bibfnamefont{M.~H.} \bibnamefont{Devoret}}
  (\bibinfo{publisher}{Plenum}, \bibinfo{year}{1992}), pp.
  \bibinfo{pages}{21--108}.

\bibitem[{\citenamefont{Averin and Nazarov}(1990)}]{AverinNazarov}
\bibinfo{author}{\bibfnamefont{D.~V.} \bibnamefont{Averin}} \bibnamefont{and}
  \bibinfo{author}{\bibfnamefont{Y.~V.} \bibnamefont{Nazarov}},
  \bibinfo{journal}{Phys. Rev. Lett.} \textbf{\bibinfo{volume}{65}},
  \bibinfo{pages}{2446} (\bibinfo{year}{1990}).

\bibitem[{\citenamefont{Geerlings et~al.}(1990)\citenamefont{Geerlings, Averin,
  and Mooij}}]{Geerlings}
\bibinfo{author}{\bibfnamefont{L.~J.} \bibnamefont{Geerlings}},
  \bibinfo{author}{\bibfnamefont{D.~V.} \bibnamefont{Averin}},
  \bibnamefont{and} \bibinfo{author}{\bibfnamefont{J.~E.} \bibnamefont{Mooij}},
  \bibinfo{journal}{Physical Review Letters} \textbf{\bibinfo{volume}{65}},
  \bibinfo{pages}{3037} (\bibinfo{year}{1990}).

\bibitem[{\citenamefont{Eiles et~al.}(1992)\citenamefont{Eiles, Zimmerli,
  Jensen, and Martinis}}]{Eiles}
\bibinfo{author}{\bibfnamefont{T.~M.} \bibnamefont{Eiles}},
  \bibinfo{author}{\bibfnamefont{G.}~\bibnamefont{Zimmerli}},
  \bibinfo{author}{\bibfnamefont{H.~D.} \bibnamefont{Jensen}},
  \bibnamefont{and} \bibinfo{author}{\bibfnamefont{J.~M.}
  \bibnamefont{Martinis}}, \bibinfo{journal}{Phys. Rev. Lett.}
  \textbf{\bibinfo{volume}{69}}, \bibinfo{pages}{148} (\bibinfo{year}{1992}).

\bibitem[{\citenamefont{Paul et~al.}(1994)\citenamefont{Paul, Cleaver, Ahmed,
  and Whall}}]{Paul}
\bibinfo{author}{\bibfnamefont{D.~J.} \bibnamefont{Paul}},
  \bibinfo{author}{\bibfnamefont{J.~R.~A.} \bibnamefont{Cleaver}},
  \bibinfo{author}{\bibfnamefont{H.}~\bibnamefont{Ahmed}}, \bibnamefont{and}
  \bibinfo{author}{\bibfnamefont{T.~E.} \bibnamefont{Whall}},
  \bibinfo{journal}{Phys. Rev. B} \textbf{\bibinfo{volume}{49}},
  \bibinfo{pages}{16514} (\bibinfo{year}{1994}).

\bibitem[{\citenamefont{Pasquier et~al.}(1993)\citenamefont{Pasquier, Meirav,
  Williams, and Glattli}}]{Pasquier}
\bibinfo{author}{\bibfnamefont{C.}~\bibnamefont{Pasquier}},
  \bibinfo{author}{\bibfnamefont{U.}~\bibnamefont{Meirav}},
  \bibinfo{author}{\bibfnamefont{F.~I.~B.} \bibnamefont{Williams}},
  \bibnamefont{and} \bibinfo{author}{\bibfnamefont{D.~C.}
  \bibnamefont{Glattli}}, \bibinfo{journal}{Physical Review Letters}
  \textbf{\bibinfo{volume}{70}}, \bibinfo{pages}{69} (\bibinfo{year}{1993}).

\bibitem[{\citenamefont{Covington et~al.}(2000)\citenamefont{Covington, Keller,
  Kautz, and Martininis}}]{Covington}
\bibinfo{author}{\bibfnamefont{M.}~\bibnamefont{Covington}},
  \bibinfo{author}{\bibfnamefont{M.~W.} \bibnamefont{Keller}},
  \bibinfo{author}{\bibfnamefont{R.~L.} \bibnamefont{Kautz}}, \bibnamefont{and}
  \bibinfo{author}{\bibfnamefont{J.~M.} \bibnamefont{Martininis}},
  \bibinfo{journal}{Phys. Rev. Lett.} \textbf{\bibinfo{volume}{84}},
  \bibinfo{pages}{5192} (\bibinfo{year}{2000}).

\bibitem[{\citenamefont{Flensberg}(1997)}]{Flensberg}
\bibinfo{author}{\bibfnamefont{K.}~\bibnamefont{Flensberg}},
  \bibinfo{journal}{Phys. Rev. B} \textbf{\bibinfo{volume}{55}},
  \bibinfo{pages}{13118} (\bibinfo{year}{1997}).

\bibitem[{\citenamefont{Averin and Odintsov}(1989)}]{AverinOdintsov}
\bibinfo{author}{\bibfnamefont{D.~V.} \bibnamefont{Averin}} \bibnamefont{and}
  \bibinfo{author}{\bibfnamefont{A.~A.} \bibnamefont{Odintsov}},
  \bibinfo{journal}{Physics Letters A} \textbf{\bibinfo{volume}{140}},
  \bibinfo{pages}{251} (\bibinfo{year}{1989}).

\bibitem[{\citenamefont{Zorin}(1995)}]{ZorinCoax}
\bibinfo{author}{\bibfnamefont{A.~B.} \bibnamefont{Zorin}},
  \bibinfo{journal}{Rev. Sci. Instrum.} \textbf{\bibinfo{volume}{66}},
  \bibinfo{pages}{4296} (\bibinfo{year}{1995}).

\bibitem[{\citenamefont{Wellstood et~al.}(1994)\citenamefont{Wellstood, Urbina,
  and Clarke}}]{Wellstood}
\bibinfo{author}{\bibfnamefont{F.~C.} \bibnamefont{Wellstood}},
  \bibinfo{author}{\bibfnamefont{C.}~\bibnamefont{Urbina}}, \bibnamefont{and}
  \bibinfo{author}{\bibfnamefont{J.}~\bibnamefont{Clarke}},
  \bibinfo{journal}{Phys. Rev. B} \textbf{\bibinfo{volume}{49}},
  \bibinfo{pages}{5942} (\bibinfo{year}{1994}).

\end{thebibliography}

\end{document}